\def\reuco{1}
\def\reuctw{2}
\def\reuctr{3}
\def\reucfo{4}
\def\reucfi{5}
\def\reucsi{6}
\def\reucse{7}
\def\rlro{8}
\def\rlrtw{9}
\def\rto{10}
\def\rttw{11}
\def\rttr{12}
\def\rcyco{13}
\def\rdtw{14}
\documentstyle[12pt]{article}
\begin{document}
\title{Imitation Games}
\author{
        Kunihiko Kaneko and Junji Suzuki$^{\dagger}$\\
        {\small \sl Department of Pure and Applied Sciences,}\\
        {\small \sl \phantom{}$^{\dagger}$Institute of Physics,}\\
        {\small College of Arts and Sciences}\\
        {\small \sl University of Tokyo}\\
        {\small Komaba 3-8-1, Meguro-ku, Tokyo 153, JAPAN}
}
\date{}
\maketitle
{\bf  Mutual imitation
games among artificial birds are studied.  By employing
a variety of mappings and game rules,
the evolution to the edge between chaos
and windows is universally confirmed.
Some other general features are observed,
including punctuated equilibria, and successive alternations
of dominant species with temporal complexity.
Diversity of species aided by the symbolization of artificial birds' song
are also shown.
}
%
%
%
%
\vspace{.2in}
\noindent{\bf{\S1 Introduction}}
The evolution of species in mutually interacting system has been
a problem of interests in many branches of sciences.
Is there trend to increase the complexity there?
Is there any characterization for an evolutionarily favorable state?
Determination of such a state is generally a complicated
question, involving a certain balance among several factors,
such as costs to adopt a strategy and behavior and gains by adopting it.
The recent ``dogma" for the characterization may be
``the edge of chaos' scenario.  This concept have been emphasized in
many contexts such as cellular automata, boolean networks or CML
\cite{Packard,Chris,Kauf2,KK}.
To the authors' knowledge, however, there was no clear example
presenting the evidence for evolution to the edge of chaos in
the exact sense of dynamical systems theory.
In the recent communication, we have proposed a minimal model for
the evolution to the edge of chaos, based on the dynamical
systems theory \cite{KKJS93}.
The model has been motivated by the observed complexity of bird songs:
It is known that a bird with a complex song ( with many repertoires
made from combinations of simple phrases) is stronger in defending its
territory, as is observed by Krebs with the help of loud-speaker
experiments \cite{Catch,Krebs}.
So far the reason why a complex song is evolved is unknown, although
there is a hypothesis that a complex song may give an impression of crowded
population, thus is efficient for the defense\cite{Krebs}. This
hypothesis, however, has no experimental evidences.
On the other hand, there are some reports of observations that birds try to
imitate each other's song at the defense of territory \cite{Catch}.
Combining the above two observations,
the following hypothesis has been put
forward for the explanation of such complexity of a bird song:
Birds join a battle for defense of territory through mimicking each other's
song. In other words, if a bird A can imitate bird B's song better,
then A can intrude B's territory
and can share feeds. A is advantageous in the battle for the survival.
It may be expected that a complex song is difficult to be imitated.
Hence we may expect the evolution to a complex song.
In spite of the lack of a direct experimental support
for this scenario, the concept of mutually imitating game itself deserves
much attention, as a novel problem of game theory and as
a general framework for the evolution to a complex behavior.
As an abstract game, we adopt an artificial bird (or player) whose song is
generated by a nonlinear map.
Birds play imitation games with each other.
We assume that the winner of the game is
fitter in the survival, which is taken into account
by the population dynamics as a growth rate or
by a simple replacement of losers by
winners. Mutational changes of parameters for the song dynamics
are introduced in the course of replication
(See for details the modeling in section 2).
We have to emphasize two advantages in our modeling here.
\begin{itemize}
\item Since our artificial birds generate their songs by nonlinear mapping,
   the songs have complexity with real numbers.  In most game theory,
   strategies are chosen from discrete sets, while the strategies (songs)
   here are chosen from continuously many sets.

\item Since our parameters have a clear meaning in the nonlinear mapping,
   the meaning of the edge of chaos is quite clear in the exact sense of
   dynamical systems theory.
\end{itemize}
Since all players are labeled by continuous parameters,
no clear speciation is given in advance.
Judging from the result of simulations, however, it seems that
we are entitled to divide players into several species.
With high mutation rates, the coexistence of
several species has been observed.  With the use of much smaller
mutation rates,
we have explicitly shown the evolution to the edge of chaos, in the
exact sense of nonlinear dynamics.
In addition, the evolution occurs towards the borderline
between a window and chaos, not to the first onset of chaos.
This observation leads to the hypothesis of the evolution to
the border between observable and invisible chaos.
Some other features emerge through the evolution, such as
punctuated equilibria and abrupt, complex alternations of
dominant species.
In the previous report \cite{KKJS93}, we have adopted
\begin{itemize}
\item the logistic map as a song generator
\item euclidean criterion for imitation
\item ``infinite dimensional" lattice (all-to-all game)
\item  (see the meanings of these terms in section 2).
\end{itemize}
Here we report studies on
several variants of the model, to test the universality of the results
obtained previously.
For example, we put birds on a 2D square lattice, allowing only for
fights among the players on the nearest neighbor sites.
In most of cases, the results in the previous paper \cite{KKJS93}
are reproduced,
supporting the universality of the concept of the ``edge of chaos",
and the edge of windows.
Furthermore, we find several exotic phenomena such as the diversity
(or coexistence) of species assisted by discrete symbolization
of birds' songs.
This paper is organized as follows.
In the next section, we present details of our modeling.
Several factors defining our models are described respectively.
In section 3 we survey the results of the model
with ``Euclidean criterion",
parts of which were also reported previously \cite{KKJS93}.
These results provide basic notions and framework for
later sections.  Characteristic features in the
imitation games are shown there.
We will give some intuitive reasons why the system evolves to the
edge between chaos and a window. Moreover,
it will be shown that our ``edge of chaos"
 scenario is stable against choices of
the topology of players. In section 4, results with the choice
of symbolized criterion, i.e., ``LR criterion" are reported.
Such symbolization of dynamics is shown to
lead to the increase of fluctuation, resulting the diversity of
species.
In section 5, we revisit the problem of the priority of the edge between
chaos and a window, by changing
the birds' song generator (to the tent or the circle map).
In particular,
we will confirm the priority by simulating a system
with the coexistence of two distinct types of song
mappings with and without windows
(i.e., the logistic and tent maps).
In section 6, we will discuss complex dynamics
of the imitation games by  allowing for different dynamics
for the ``song generation" and imitation processes.
Summary and discussions are given in section 7.
%
%
%
%
\vspace{.2in}
\noindent{\bf {\S 2 modeling}}
Our model consists of the three stages,
the song generation process, 2-persons' imitation
game and reproduction.
We will discuss these three procedures separately.
%
%
\noindent{\bf 1: Song generation (Song dynamics)}:
 As a ``song", we use a time series generated by a nonlinear map.
In the previous report \cite{KKJS93},  the logistic map:
\begin{equation}
x_{n+1}=1-a x_n^2
\end{equation}
is adopted for this purpose,
since it is one of the most investigated maps
as a standard model in nonlinear physics.
A bird, say, the i-th bird, possesses its own parameter
$a(i)$. We regard time series $(x_0(i), x_1(i), \cdots)$
generated by the map as the i-th bird's song.
For most examples of simulations here,
the logistic map is again adopted, while the
other maps for a bird's song generator
are also adopted to test the universality of our results.
Here simulations with the use of
the tent map or the circle map will be given in \S 5.

%
%
\noindent{\bf 2: 2 person imitation game}:
Three factors are involved in defining the 2-persons' game;
topology for players, imitation , and game processes (or
criterion for winners).  Here we
give detailed explanation of each process separately.
\noindent{\sl  2.a) topology for players}
In the previous report \cite{KKJS93},  players are assumed to fight against
all other players.  In the sense of statistical physics,
players live on an infinite dimensional lattice. Any spatial information
is not included in this modeling.
Here we also investigate cases where  players are put on a
two-dimensional square lattice with the periodic boundary
condition.  For most simulations here,
we adopt the lattice of $ 30 \times 30$.
\noindent{\sl 2.b) imitation process: choice of an initial value for an
imitating song:}
Each bird player $i$ has to choose an initial condition, so that
the time series by its own dynamics  can
imitate the other player's song better.  Here we use the following imitation
process for simplicity.
For given transient time steps $T_{imi}$, a player
(mimicker) modifies its dynamics
with a feedback from the other player(singer):
\begin{equation}
x_{n+1}(1)=f_1 [(1-\epsilon)x_n (1)+\epsilon x_n (2)]
\end{equation}
By this dynamics, the player 1 adjusts its value $x_n(1)$ by referring to the
other player's value ($x_n(2)$).  Here $\epsilon $ is a coupling parameter
for imitation process.  After repeating this imitation process
for $T_{imi}$ steps, the player 1 uses its own dynamics
$x_{n+1}(1)=f_1 (x_n(1))$.  In other words, the above process is used
as a choice of the initial condition for the imitation of the
other player's dynamics.  The coupling parameter $\epsilon $ also varies by
players.
However, the distribution of this parameter seems ``irrelevant", as
far as we judge from our numerical results.
Thus we will skip the discussion on this parameter hereafter.

\noindent{\sl 2.c): Game}:
After the player $1$ completes the above imitation process,
2 players $1$ and $2$ are decoupled and
generate songs by their own dynamics.
Then we calculate a quantity $D(1,2)$ measuring
the distance between
the imitating time series $\{ x_n (1) \}$, and the singing one
$\{ x_n (2) \}$.
By changing the role of two players, $D(2,1)$ is also measured.
If $D(1,2)$ is smaller than $D(2,1)$ the player $1$ imitates better
the other's song and wins this 2-persons' game ( and vice versa).
Definition of the distance $D(i,j)$ gives a criterion
for the winner and may be crucial in the game.
Here we adopt either of the following two criterions.
\noindent{\it 2.c.1) Euclidean distance}:
As the measure, we choose $D(1,2)=\sum_{m=1}^{T} |x_m (1)-x_m (2)|^2$ over
certain time steps $T$.  This choice is just a normal
Euclidean distance.  (
Of course, the distance should be cordial for the circle map,
since $x$ is on a circle.)
\noindent{\it 2.c.2) LR-code}:

It is often useful to symbolize the time series
to some discrete sets rather than to use the real number $x$.
The time series of the logistic map,  for example, can be traced
by 2 symbols, represented by R ($x>0$) and L ($x<0$) \cite{logistic}.
Using such symbolization, the distance between two trajectories is
measured by the number of unmatched symbols over some time steps.
In the above symbolization for the logistic map, the distance
is thus defined by
\begin{equation}
D(1,2)=\sum_{m=1}^T IS(x_m(1),x_m(2))
\end{equation}
 with the notation
$IS(a,b)=1$ if $ab<0$ and 0 otherwise.
\noindent{\bf 3 Reproduction}

Fitness of the players are given by
the results of games.
Offsprings are reproduced according to the results of fights.
We adopt either of the following two for this process :
\noindent{\sl 3.a) Replacement}:
This might be the simplest way; the parameters of losers are replaced
by those of winners at every game. We adopt this replacement rule
for players on a 2D lattice.
\noindent{\sl 3.b) Score}:

After each game, the winner gets
a point $W$, while a loser gets L ($W>L$)  (Both get $(W+L)/2$ in the
case of draw.)  After iterating a large number of games, the population
distribution is updated in proportional to the score of  players,
with the further restriction that the total population be constant.
At both of these reproducing stages, we  include
mutational errors to the parameters;   The parameters $a$
and $\epsilon$ are changed to
$a+ \delta$, $\epsilon + \delta '$, where  variables $\delta$ and
$\delta '$ are random numbers chosen from a suitable distribution
( we use a homogeneous random distribution over $[-\mu , \mu]$ or Lorenzian
distribution ($P(\delta )= 1 /\{ \mu (1+ (\delta /\mu )^2)\} $).  The latter
choice is often useful, since the former choice inhibits a large jump of
parameters  and often the parameter values are trapped at intermediate values,
 while the latter can provide a wider possibility of ``species".

These three processes define our model. Because of the numbers of variations,
we will sometimes specify a model by sets of rules such as
[lattice, logistic, LR, Score, $5 e-4$],
which means that players are on a 2D lattice, and generate songs by the
logistic map,  that the distance
between two songs is measured by LR symbolic sequence, and that the
population is updated in proportional to player's score with the
mutation rate $\mu =5 \times 10^{-4}$.

Throughout this paper time step is measured in the unit of the total number of
players $N$.  In the lattice case one step is defined as the time
necessary for one game for each player, i.e., totally
$N$ games for $N$ players.  In the all-to-all game, it is
defined as the time required that each player plays a game for all
othe players ( i.e., totally $N(N-1)/2$ games for all players).
%
%
%
%
\vspace{.2in}
\noindent{\bf \S 3 Basic Results: Simulations with Euclidean Criterion}
Let us start our presentation by giving numerical results with the
Euclidean criterion for distance.
First, we present a result of the game with high mutation
parameters. The model is defined by [lattice, logistic, euclid,
replacement, ?]. In figure[\reuco], we present the distribution of the
logistic parameter $a$ after 10 time steps. It clearly shows
the coexistence of two species; one is around $a=0.75$ and the other is
around $a=1.75$. The former corresponds to the bifurcation point (period
1 $\rightarrow$ period 2) while the latter to the edge between
the period 3 window and chaos.
The snapshot configuration is depicted in Fig[\reuctw], where dark dots
mean players with $ a \sim 0.75$ while bright dots give those
with $ a \sim 1.75$. This example already shows the typical
character of advantageous states for some bifurcation points
( in other words, ``edge of something").

To see the fitness by players we switch to use
a model with all-to-all games [infinite, logistic, euclid, score, ?],
presented in the previous communication, since it deserves a
prototype of our models.
The average score of players is plotted as
a function of the bifurcation parameter $a$
by taking a high mutation rate, to allow for
the existence of players with a wide range of parameters $a$[Fig \reuctr].
We can see many peaks, which correspond to
the bifurcation point from period-2 to period-4,
and from period-4 to period-8, and the edge of period-3 window,
period-5 window, period-4 window, and so forth.  This score landscape is
rugged as is often discussed in spin-glass type models in biology
\cite{landscape,SG}.
Here this landscape is not implemented as a model itself
(like energy function in spinglass type models \cite{SG}),
but is emergent through the
evolution.  Indeed this landscape depends on the population
distribution at the moment.
To see the advantage of ``the edge of chaos" in the score,
we have plotted the score of birds as a function of the
Lyapunov exponent $\lambda $ [Fig \reucfo].
As is shown there, the score has a clear
peak around $\lambda =0$.  Indeed this peak is
supplied at the edge of period-3 window and chaos ($a \approx 1.75$),
and the edge of period-4 window and chaos ($a \approx 1.94$).

A high mutation rate, however, leads to a large fluctuation,
which makes it difficult to identify each ``species" precisely.
Let us therefore concentrate on low mutation rate cases for the
rest part of the present paper.
In Fig.[\reucfi], temporal evolution of the
average of the parameter $a$ over all players is plotted.
Plateaus are observed successively,
providing an explicit example for punctuated equilibrium \cite{punceq}.
At the temporal domain with a plateau, the deviation from the average
value is typically very small.  Each plateau corresponds to the
bifurcation points and the edges of chaos mentioned at
the score landscape.  Finally, the birds reach the parameter
for the edge of period-4 window ($a \approx 1.94$), and stay there
\footnotemark.
\footnotetext{
The averaged score plotted as a function of the Lyapunov exponent $\lambda$
has a broad peak at $\lambda =0$, which extends to the region $\lambda <0$
rather broadly.  The score has a sharp drop at the side of
$\lambda >0$, on the other hand.
With the increase of $T$, the peak slightly shifts to
$\lambda>0$ while drops at $\lambda <0$ get sharper.}

Thus we have seen the priority of ``the
edge of chaos", and the evolution to it.
To be more precise, we have to note
that our edge state lies between a window and
chaos.  At windows, the logistic map can provide
chaotic transients before the dynamics settles down to a
stable cycle.  Thus the existence of transient chaos should be useful to
imitate a dynamics of different nature.
A window at a higher nonlinearity regime includes a variety of unstable
cycles, as coded by Sharkovskii's ordering \cite{logistic}.
Therefore it can provide a larger variety of dynamics, as transients.
This might be the reason why the edge of windows is strong in our
imitation game.
The above speculation suggests the importance of transient chaos,
besides the edge of it, for the adaptation to  a wide range of external
dynamics.
Let us incorporate again the lattice structure, by taking the model
[lattice,logistic, euclid, score, ?].
Temporal evolution of the average logistic parameter value $a$ is
plotted in Fig[\reucsi].
Again we see the evolution to the edge of chaos (windows) after punctuated
equilibria.
Thus the previous model with all-to-all game is worth of a good
mean field theory to our problem.
The lattice structure  enhances the fluctuation
as is seen by comparing the flatness of plateaus
in Figs. (5) and (6) or noticing the lack of some plateaus here.
To get some spatial information,  we measure the distribution of
the maximum connected cluster sizes of a species,
since admissible parameter regions are separated enough to
be regarded as distincted species.
We have plotted the temporal changes of the maximal cluster sizes
around the time steps for
abrupt changes of the mean logistic parameter [Fig \reucse].
Successive transitions of dominant species are clearly shown.
An interesting question here is if the lattice structure helps
the coexistence of species or not.
In [Fig \reucse], both species with $a \approx 1.75$ and $a  \approx 1.86$
increase their populations together, which seems to suggest a kind
of symbiosis.
This example, however, is not so decisive to claim the spatial differentiation
supporting  the coexistence of species.  Further studies are necessary.
The examples in the present section
show that the ``edge of chaos" concept is valid irrespective
of the topology of players. We stress again the significant role of the
windows, which will be further addressed in section 5.
%
%
%
%
\vspace{.2in}
\noindent{\bf \S 4 Symbolization induced complexity and diversity; LR
criterion}:
Next we study the case with the use of the distance
in terms of symbolic codes. The motivation for this choice is the
examination of the effect of symbolization
in communication codes:  Does the symbolization lead to
increase or decrease of diversity and complexity?
The temporary evolution of the average logistic parameter
is given in Fig[\rlro].
In this case, advantageous players have such parameters that
the symbolic sequence changes its pattern there, e.g., $a=1,1.3, \cdots$.
It is not difficult to see that the dominant species
lie again at window values in long intervals.
Clearly, this example shows the stability of ``edge between chaos and window"
scenario against the choice of criterions in  the 2-persons' game.
Judging from the amplitudes in Fig[\reucse] and [\rlro], we conclude that
the fluctuation here is further enhanced from the Euclidean
criterion case.
In the lattice version, the average
values of $a$ go up and down temporally with some
transient ``disordered" intervals.  No ``final" parameter ( such as
$a \approx 1.94$ which is robust and is reached from any distributions
in the Euclid criterion) exists here.  The parameters switch (almost)
forever.  After staying
at the edge between some window and chaos, the parameter switches to
the edge of another bifurcation point or to an edge.
In the example of Fig[\rlro], successive transitions between
$a \approx 1.75$ (period 3) and $a \approx 1.94$ (period 4) are
observed.  Such complex alternations between ``ordered states"
are typically seen in the chaotic itinerancy \cite{CI}, although the
present switching process may be
of stochastic nature (not of deterministic).
Moreover, coexistence of different species (parameters) in the LR case
is strongly enhanced compared with the Euclidean case with
the same mutation rate.
See Fig.[\rlrtw] for snapshots of the distribution of players
with respect to distinct sets of the logistic parameter $a$.
The diversity is enhanced in
the symbolic criterion.
The increase of diversity is also seen in simulations with the
all-to-all game.  Roughly speaking, players with  $a \approx 1.75 $
(period-3 window) are rather robust and have occupied a large ratio of
population. ( As the fluctuation is increased, a window with a larger
interval (such as the period-3 one)
is more robust, which is also true of a model with a high mutation rate
with Euclidean criterion \cite{KKJS93}).
Still, we have seen coexistence of other parameters
such as $a \approx 1.94$ besides the above group.
When the average is plotted with time, it fluctuates around
$a \approx 1.75 \sim 1.78$, while the variance of the fluctuation
remains large even in a very small mutation rate regime ( say $10^{-4}$).
Summing up the present section, we have found that
the symbolization enhances complexity of dynamics and diversity of
species.  We have found coexistence of two groups, and successive
switches of dominant species for ever.
What is the origin of this diversity and complexity?
One possible guess is as follows.
By the binary symbolization,  players with a certain range
of the logistic parameter come to possess an identical temporal pattern in
singing, whereas fine structures within the parameter
interval are distinguished in the Euclidean (or
analogue) measure.  Thus a ``strong" pattern of time series
which was observed only in some very narrow window regime
may be stabilized and exist in a wider parameter regime.
The species, for example, at $a \sim 1.63$,
can have a larger chance for the survival.
By the above mechanism of symbolization,
effective strength of players is further averaged out.
Difference between two songs in the game is much weaker than
the Euclidean one  \footnotemark.
Relationships between two players with different edges are more subtle,
and no robust parameter exists.
The switching dynamics and the coexistence may reflect this averaging
effect by the symbolization.
\footnotetext{ Another way of weakening
the difference (but keeping ananlogue criterion)
is the use of modified definition for distance measure;
$D(1,2)=\sum_{m=1}^{T} |x_m (1)-x_m (2)|^y$ with $y <2$.
We have studied the case $y=1$, which also
enhances the fluctuation and instability of dominance of a species,
similarly with the LR criterion.  Such modification may be rather
superficial, however,  as is immediately seen by considering the
extreme limit $y \sim 0$.}
If this scenario of diversity is universal, it suggests the
importance of symbolization or discretization for the diversity.
Why is the human language so diverse? Is the diversity caused by our
symbolization ability from analogue vocal signals?
%
%
%
%
\vspace{.2in}
\noindent{\bf \S 5 Universality of the priority of the edge between window and
chaos}
Here we examine the universality of the scenario in \S 3,
by choosing a few different mappings for the song dynamics.
In this section, we will simulate using the tent map
and the circle map as the song generator. The results support our hypothesis
on the  priority of the windows, discussed in \S 3.

\noindent{\sl Tent Map}
We choose the following tent map parameterized by one parameter,
\begin{equation}
x_{n+1}=a(\frac{1}{2}-|\frac{1}{2}-x_n|),
\end{equation}
as the song dynamics.
Thus each bird possesses two parameters $a(i), \epsilon(i)$ as in the case
of the game with the logistic map.
In the tent map, there is no window structure.  With the increase of $a$,
chaos appears at $a=1$, which is the only edge of chaos.
For $a>1$, no bifurcation structure exists.
Numerical results of the evolution of the imitation game show
that the average value of the parameter $a$ evolves to 1,
the onset of chaos for the tent map [Fig \rto, \rttw],
irrespective of criterions.  Since $a=1$ is the only edge,
this result is expected.
For the logistic map, we have given a plausible
explanation to the priority of windows  in section 3.
Is this edge of window stronger than the above onset of chaos in the
tent map?
Since no window exists in the tent map, it is interesting
to study a system where birds with the logistic map and with
the tent map can coexist and compete. Here
we slightly modify the logistic map to
\begin{equation}
x_{n+1}=2a x_n(1-x_n)
\end{equation}
so that the variables (parameter $a$) in both maps take values in a
same range [0,1] ($a $[0,2]).
We put players on a 2D square lattice, and adopt the symbolic
criterion (L for $x<.5$ and R for $x>.5$).
Initially,  a bird's song takes either logistic or tent map
randomly with equal weights.  Starting from the population with
small $a$ values,
the mean value of $a$ for logistic birds evolves to a punctuated
equilibrium value $a \sim 1$ after few steps. (See Fig.[\rttr]).
At this equilibrium
state, the ratio of the birds with the logistic map to  those
with the tent maps is kept almost constant $ \sim 8:1$.
Then the logistic parameter shows the abrupt change to the value corresponding
to the period-3 window (or the period-4 window, depending on system size).
During the change, the birds with
tent maps are completely terminated, which means that the onset of chaos
in the tent map is defeated by the windows in the logistic map.
This result may deserve a support to our hypothesis on the priority
of windows' edge.
\noindent{\sl Circle map}
We have also studied the case with the use of the circle map
\begin{equation}
x_{n+1}=x_{n}+a \sin(2 \pi x_n) +d  \qquad {\hbox {mod} }1
\end{equation}
as the song generator \cite{KK-book}.
Now each bird possesses three parameters $a(i), d(i), \epsilon(i)$.
We determine the rule for 2 persons' game by
measuring the cordial distance between two trajectories of maps as
noted previously.
Numerical results again show abrupt changes of parameters
with successive punctuated equilibria.
Synchronized temporal motions between average values of $a$ and
$d$ are clearly seen in Fig.[\rcyco]. The average of $d$ seems to take an
``equilibrium" value either $ \sim 0$ or  $ \sim 0.5$, resulting a shift
of the origin of variables $\{ x_{n}(i) \}$. On the other hand,
average $a$ lies again at window values (to be
precise, those with $d \sim 0$) $a \sim -2, -1.5, 1.5$ etc. \footnotemark
\footnotetext{
To see the role of $d$, we also have carried out some simulations
fixing $d=0$ and
with same choices for other parameters. In this case, the magnitude
of average $a$ seems monotonously increasing with punctuated equilibria
at windows $\sim -5, -7 \cdots$.
 Thus $d$ is not irrelevant,
but plays a role in suppressing the magnitude of the temporal variation of
$a$. }
Summarizing this section, the significance of the window
structures in the imitation game, discussed in section 3, is further
confirmed by changing the generator for birds' songs.
%
%
%
%
\vspace{.2in}
\noindent{\bf \S 6  Dual dynamics}
Before closing this report, let us present an example stressing
clearly an aspect of dynamical complexity in the imitation game.
Let the logistic map be again the song generator.
This time, however, we assign two parameter values for the logistic map
to each player so that it can use a different value in imitating and in
singing.  This is not an irrelevant complication of the model.
Rather, there is no a priori reason to believe that a bird should
use an identical parameter for the two processes.
For imitation games with one parameter, the restriction of using
an identical parameter in singing and mimicking may be a
cause to achieve an optimal value of either singing or mimicking, and
lead to an edge.
In the two parameter game it may be possible
that each process evolves to its own optimal value.
Then one might expect that the results of the two parameter games' would
be either of the following:
I. the parameters in singing and imitating evolves to the same
``edge of chaos" value; or II. the parameter in singing reaches
a value for a fully chaotic state $(\sim 2.0)$ while
the imitating value also evolves to
a high nonlinear region.
Surprisingly, the observed result supports neither of these.
In the Euclid criterion, the singing parameter again stays around
$a \approx 1.94$ the edge of window of period 4, while
the imitating parameter stays around some other windows' edge
( e.g., $a \approx 1.75$) or other bifurcation points ( e.g., $a\approx 0.75$,
bifurcation point from period-1 to 2), depending on the mutation rate
and the topology of the game.
Results with symbolic criterion are much more complex.
Let us take an example with [lattice, LR, 5e-5].
The singing parameter switches among $a \sim 1.77,1.94,2$, in synchronization
with the changes of imitating parameter $a \sim 1.1,1.3,1.77$(Fig.[\rdtw]).
This switch continues for ever as far as we have checked.
The above result may be interpreted as follows:
Optimal values for singing and imitating
parameters are not unique, but there may be several local minima.
Thus optimal values for the singing parameter
depend strongly on the population distribution of the mimicking parameter
at that time, and vice versa.
A small disturbance is enough to change both the optimal values
and the population distribution completely.
Synchronized changes between the two parameters
reflect such a subtle balance.
As stressed in section 3, the fitness function is not
given a priori but is emergent through the evolution.
For one-parameter games, however, the landscape is
rather stable  once the system evolves to the ``edge of chaos".
In the dual dynamics game, the landscape, besides its
ruggedness, varies in a quite complex manner.
Thus the model provides an illustration of a novel problem:
evolution of a system with a dynamically changing rugged
landscape.
Summing up the present section,
the evolution to the edge between chaos and windows is
still valid in dual dynamics but the population dynamics is more
complicated with successive alternations of dominant species, and
synchronized changes between the two parameters.
This example illustrates that
static (or ``equilibrium"  ) characterizations are not enough
for the imitation games, but
the understanding of dynamics
is essential to them.
%
%
%
%
\vspace{.2in}
\noindent{\bf \S 7 Summary and Discussions}
In this report, we have examined the evolution of imitation
games.  It is shown that the evolution to the ``edge of chaos"
is universally observed irrespective of topology of players,
criterions for winners in a fight, and the choice of the map as a
song generator.
Besides the edge of chaos, we have also given a support to the priority of
the edge of windows.   It is clearly demonstrated by
the dominance of the window's edge of the logistic map over
the onset of chaos in logistic and tent maps.
Though birds with both maps
can coexist with parameter values around the onset of chaos,
the players with the tent map  are completely terminated with the abrupt
change of the logistic parameter to the period-3 window's value.
Thus we propose an additional scenario here; {\bf
the evolution towards the edge between chaos and window}.
This edge corresponds to the border between observable
and invisible chaos.  At a window value, there is
topological chaos in dynamics.  Although the final attractor is a periodic
cycle for almost all initial conditions with a probability measure 1,
there are chaotic
orbits from non-measurable initial conditions (on a Cantor set).  In
connection with this
topological chaos, there are chaotic transients before an orbit is attracted
to the periodic cycle.
The existence of topological chaos assures
a variety of unstable periodic orbits.
Thus a player with a window parameter
can imitate a large variety of periodic orbits, showing its dominance over
players with
periodic dynamics without topological chaos ($a<1.4011\cdots$ in our logistic
map).  By the transient chaos, the dynamics has an ability to imitate
chaotic time series roughly up to
the length of transients, which diverges at the edge.
As a song generator, dynamics at a window parameter
can provide a large variety of orbits as
transients.  Since the transient length diverges at the edge,
a high variety of songs is maintained.  Thus generated songs
are not easily imitated by periodic or chaotic dynamics.
Summing up, our {\bf window's edge} scenario is based on
the ability of creation of complexity with topological and
transient chaos.  Transient chaos has a potentiality to
adapt a wide range of external dynamics, while the orbit at the attractor
is not complicated.  The transient chaos may be important in
a wide area of biological information processing,
and our {\bf window's edge} scenario may be applied universally in
the evolution and adaptation.
Besides the characterization of the evolutionary advantageous states,
we have discussed the origin and maintenance of diversity of species,
in a symbolized criterion and in dual dynamics for songs and mimicry.
If there is a given fixed fitness landscape,
it would be reasonable to expect that the fittest species would
dominate the world. Indeed, without any mutations or changes of environment
the diversity of species may not be sustained.
There can be several possibilities in the origin of diversity.
A simple answer to the origin is the spatial differentiation mechanism.
Indeed our simulation shows coexistence of some species
in a 2-dimensional topology.
The diversity is increased with the use of spatial game.
However, a more interesting discovery in the present paper
is {\bf symbolization induced diversity}.  By adopting a symbolized
criterion, the diversity is enhanced drastically.
At some stage of information processing in the brain,
symbolization is often adopted.  Thus we may expect that
the symbolization induced complexity can be one origin of diversity
in signal and language.   \footnotemark
\footnotetext{ Did God tell us the symbolization to destroy the Tower of
Babel?}
In the symbolized criterion, the population dynamics is also complicated.
It shows successive punctuated equilibria for ever.
The complexity is further enhanced with the choice of dual dynamics
for songs and imitations.
A variety of window's edge states appear successively, providing
temporal complexity.  In a large system size,
this complexity is easily expected to lead to spatial diversity,
since the successive changes of dominant species cannot
be synchronized over all lattice regions.
The diversity induced by temporal chaos is
recently discussed as homeochaos scenario \cite{homeo},
where the population dynamics of
species shows weak and high-dimensional chaos.
Our window's edge scenario shares the notion of ``weak chaos" with the
homeochaos scenario.
In the homeochaos, we have also seen successive switches among
ordered states, noted as chaotic itinerancy \cite{homeo2,clust}.
Detailed study of dynamical mechanisms of the successive switching in our
case will be necessary in future.
Our imitation game provides a universal route to the evolution
to complexity.
Evolution pressure of complexity to escape from imitation
can be also conceptually used in
many examples in the biological or social evolution, besides our original
motivation to a bird song.  Such examples
may include the evolution of a communication code
( "secret code"), Batesian mimicry \cite{Maynard},
and social structure \cite{Hubler}.
\vspace{.2in}
{\em Acknowledgments}:
The authors would like to thank T. Ikegami, K.Tokita, and
S. Adachi for useful discussions. The work is partially supported by
Grant-in-Aids for Scientific
Research from the Ministry of Education, Science, and Culture
of Japan.

\pagebreak
Figure Caption
\vspace{.1in}
Figure 1   Snapshot distribution of the logistic parameter of the players
after 10 time steps. Mutation parameter $\mu$ is chosen
to be 0.001, and lattice size is $30 \times 30$.
Two peaks corresponding special points for the logistic
map are observed (see text).  Unless otherwise mentioned,
we set $T_{imi}=T=30$ throughout the present paper.
\vspace{.1in}
Figure 2   Snapshot pattern of the
configuration of distribution of
the logistic parameter, corresponding to Fig.1.
Here a bright (dark) pixel corresponds to
a player possessing $a \sim 1.75 (0.77)$, respectively.
\vspace{.1in}
Figure 3
Emergent landscape: Average score for the players with
parameters within $[a_i, a_i +\Delta]$ is plotted
for $a_i=-1+i\times \Delta$, with the bin size $\Delta =0.001$.
We have adopted $W=10,L=1$, $T_{imi}=255$,
and $T=32$.  Simulation is carried out with the rule
[infinite,logistic,Euclid,score,0.1].
starting from the initial parameter $a=0.6$ and $\epsilon =.1$.
Sampled for time steps from 1000 to 1500, over all players ( whose number is
fixed at 200).  (adapted from \cite{KKJS93}).
\vspace{.1in}
Figure 4
Average score of the game vs. Lyapunov exponents.
Simulation is carried out with the rule
[infinite, logistic, Euclid, score, 0.02],
$T_{imi}=255$ and $T=32$. We have adopted $W=10,L=1$.
starting from the initial parameter $a=0.9$ and $\epsilon =.1$.
Average scores are obtained from the
histogram of  Lyapunov exponents, for which we use a bin
size of 0.01 for $-1<\lambda<1$,
while it is set at 0.1 for $\lambda<-1$ ( since the sample there
is rather sparse).
Sampled over
time steps from 500 to 750 over all players ( whose number is
fixed at 200).
\vspace{.1in}
Figure 5
Two examples of the temporal evolution
the average parameter $a$:
Simulation is carried out with the rule
[infinite, logistic, Euclid, score], and
with mutation rates $5 \times 10^{-4}$ and $8 \times 10^{-4}$.
$T_{imi}=200$ and $T=50$.
starting from the initial parameter $a=0.6$ and $\epsilon =.1$.
Average of the parameters $a$ over all players are plotted with time.
The total population is fixed at $N=200$.
\vspace{.1in}
Figure 6
Temporal evolution of the
average logistic parameter $a$.
(as is written, 1 time step = 900 2 persons' game).
[lattice, logistic, Euclid, replacement, $\mu=8.7 \times 10^{-5}$]
Initial values for $a$
are chosen randomly from $[0.6,1.1]$.
Punctuated equilibria at $a\sim 0.77$ and $1.94$ are clearly seen.
The plateau at $1.94$ continues after 3500 steps, as far as we have observed.
\vspace{.1in}
Figure 7
Temporal evolution of 4
connected clusters in the vicinity of the change
of the average logistic parameter in Fig.6. Around the time steps
[2700, 2800], two clusters ($\sim$1.74 and
$\sim 1.86$) seem to grow simultaneously.
\vspace{.1in}
Figure 8.
Temporal evolution of the
average logistic parameter $a$.
[lattice, logistic, LR, replacement, $mu=1.25 \times 10^{-4}$].
Alternations between the states
 at $a \sim 1.77$ and $\sim 1.94$
are seen.
\vspace{.1in}
Figure 9.
Successive snapshot distribution of
the logistic parameter at time steps 1000,5000,7000.
The parameters of the model are taken identical to
those in Fig.8.
Figure 10
Temporal evolution of the average tent
parameter. The rule is given by
[30$\times$ 30 lattice, tent, Euclidean, replacement,
$6.1 \times 10^{-5}$]. Initial values for
the tent parameter $a$ are chosen randomly from $[0.3,0.4]$.
The deviation from 1 is very small after establishing
the equilibrium.
\vspace{.1in}
Figure 11
Temporal evolution of the average tent
parameter with the LR criterion.
The rule is given by
[30$\times $30 lattice, tent,
LR, replacement, $8 \times 10^{-5}$].
We have also simulated smaller $\mu$ cases, but
the fluctuation remains still larger than that from the
Euclidean criterion (given in the previous Figure).
\vspace{.1in}
Figure 12
Temporal evolution of
the average value of logistic parameters $a$
(thin line) and the population of adopting a
logistic map song.
Here the lattice size is 32 $\times$ 32
(totally 1024 players) and $\mu=7.5\times 10^{-5}$.
Around the time step 4000, birds with logistic maps cover the total
lattice and those with the tent maps are terminated.
\vspace{.1in}
Figure 13
Temporal evolution of the average $a$ (thin line) and  $d$(broken
line) values.  [30$\times$ 30 lattice, circle,
Euclidean, replacement, $5 \times 10^{-5}$].
Figure 14
Temporal evolution of the
averages of the logistic paremters
for singing(broken) and imitating (thin) in dual dynamics,
with LR criterion.
The lattice size is chosen to be 30 $\times$ 30, and
$\mu=7.5\times 10^{-5}$.

\addcontentsline{toc}{section}{References}
\begin{thebibliography}{999}
\bibitem{Packard}
N.H. Packard,
in  {\sl Dynamic Patterns in Complex Systems}, eds.
J. Kelso, A.J. Mandell, M.F. Shlesinger, World Scientific, p. 293-301
(1988).
See also the paper by M. Mitchell et al. in this volume, however.
\bibitem{Chris}
C. Langton, Physica  42D (1990) 12
\bibitem{Kauf2}
S.A. Kaufmann and S. Johnson, J. Theor. Biol. 149 (1991) 467
\bibitem{KK}
K.Kaneko, Physica 34 D (1989) 1
\bibitem{KKJS93}
K. Kaneko and J. Suzuki, `` Evolution to the Edge of Chaos in Imitation Game",
in Artificial Life III (1993) eds C. Langton, in press
\bibitem{Catch}
C.K. Catchpole, {\sl Vocal communication in birds},
Edward Arnold Press, 1979
\bibitem{Krebs}
Krebs J.R., Ashcroft,R., and Weber,M, Nature 271 (1978) 539-42
\bibitem{landscape}
See e.g., S.A. Kaufmann and S. Levin, J. Theor. Biol. 128 (1987) 11.
\bibitem{SG}
M. Mezard, G. Parisi, and M.A. Virasoro eds.,
{\sl Spin Glass Theory and Beyond} (World Sci. Pub., Singapore,  1988)
\bibitem{logistic}
R. May, Nature 26 (1976) 459;
P. Collet and J.P. Eckmann, {\sl Iterated Maps on the Intervals as Dynamical
Systems}, Birkhasuer, Boston-Basel-Stuttgart, 1980.
\bibitem{punceq}
N. Eldredge and S.J. Gould,
in {\sl Models in Paleobiology}, Freeman, 1972, ed. T.J. M. Schopf.
\bibitem{CI}
K.Ikeda, K. Matsumoto, and K. Ohtsuka,
Prog. Theor. Phys. Suppl. 99 (1989) 295
I. Tsuda, ``Chaotic Neural Networks and thesaurus", in {\sl Neurocomputers and
Attention}, (eds.  A.V. Holden and
V. I. Kryukov, Manchester Univ. Press, 1990)
K. Kaneko, Phys. Rev. Lett. 63 (1989) 219;
Physica  41 D (1990) 137; 54 D (1991) 5.
\bibitem{KK-book}
K. Kaneko, ``Collapse of Tori and Genesis of Chaos in Disspative Systems",
World. Sci., 1986
\bibitem{homeo}
K. Kaneko and T. Ikegami, ``Homeochaos: Dynamics Stability of a
symbiotic network with population dynamics and evolving mutation rates"
Physica 56 D (1992) , 406-429
\bibitem{homeo2}
T. Ikegami and K. Kaneko, ``Evolution of Host-Parasitoid Network
through Homeochaotic Dynamics" Chaos 2 (1992) 397-408
\bibitem{clust}
K. Kaneko, in these proceedings
\bibitem{Maynard}
J. Maynard Smith,
{\sl Evolutionary Genetics}, Oxford Univ. Press, 1989
\bibitem{Hubler}
see A. Hubler and D. Pines, in these proceedings
for a related problem, with the use of control of chaotic dynamics
instead of our imitation.
\end{thebibliography}
\end{document}